\documentstyle{article}
\date{}
\title{Hyperbolic Orbits for Restricted Three-body Problems with Fixed Energy\footnote{Supported partially by NSF of China.}}
\author{{Donglun Wu and Shiqing Zhang}\\
{\small Yangtze Center of Mathematics and College of Mathematics, Sichuan University,}\\
{\small Chengdu 610064, People's Republic of China}}
\begin{document}
\maketitle \large \baselineskip 14pt

\begin{quote}

{\bf Abstract}\ \ The existence of hyperbolic orbits is proved for a
class of restricted three-body problems with a fixed energy by
taking limit for a sequence of periodic solutions which are obtained
by variational methods.

{\bf Keywords}\ \ Hyperbolic Orbits; Variational Methods; Restricted
Three-body Problems; Fixed Energy.

{\bf 2000 MSC:} 34C15, 34C25, 58F

\end{quote}

\section{Introduction and Main Results}

\ \ \ \ \ \ In this paper, we consider the following second order
Hamiltonian systems
\begin{equation}
   \ddot{u}(t)+\nabla V(u(t))=0\label{1}
\end{equation}
with
\begin{equation}
   \frac{1}{2}|\dot{u}(t)|^{2}+ V(u(t))=H\label{2}.
\end{equation}
where $u\in C^{2}(R^{1},R^{N})$, $V\in C^{1}(R^{N},R^{1})$.
Subsequently, $\nabla V(x)$ denotes the gradient with respect to the
$x$ variable, $(\cdot,\cdot):R^{N}\times R^{N} \rightarrow  R$\
denotes the standard Euclidean inner product in $R^{N}$ and $\mid
\cdot\mid$ is the induced norm.

The restricted three-body problem is a reduced model for the
Newtonian three-body problems. The existence of periodic orbits,
hyperbolic orbits for this model has been studied by many
mathematicians \cite{1,20,2,23,3,19,7,27,12,16} and the references
therein. In this paper, an orbit of this problem is said to be
hyperbolic if two of the three bodies remain bounded while the third
goes to infinity with vanishing velocity. A special type of
restricted three-body problem was considered by Sitninkov \cite{25}
and Moser \cite{26}: Under Newton's law of attraction, two mass
points of equal mass $m_{1} = m_{2} = \frac{1}{2}$ moving in the
plane of their elliptic orbits such that the center of masses is at
rest, the third mass point $m$ which does not influence the motion
of the first two moving on the line perpendicular to the plane
containing the first two mass points and going through the center of
mass. Let $u$ be the coordinate describing the motion of $m$ and the
center of mass of the first two mass points is at the origin. The
restricted three-body problem consists in determining $u$ such that:
\begin{eqnarray*}
 -\ddot{u}(t)=\frac{u(t)}{(|u(t)|^{2}+|r(t)|^{2})^{3/2}},
\end{eqnarray*}
where $r(t)=r(t+2\pi)$ is the distance from the center of mass to
one of the first two mass points. For a small $\varepsilon>0$, the
function $r$ has the form (see Moser \cite{26}):
\begin{eqnarray*}
 r(t)=\frac{1}{2}(1-\varepsilon\cos t)+O(\varepsilon^{2}).
\end{eqnarray*}

Souissi \cite{27} used variational minimax methods and
approximations to prove the existence of at least one parabolic
orbit of the circular restricted three-body problem with
$0<\alpha<1$ for
\begin{eqnarray}
 \ddot{u}(t)+\frac{\alpha
 u(t)}{(|u(t)|^{2}+|r(t)|^{2})^{\frac{\alpha+2}{2}}}=0.\label{6}
\end{eqnarray}

With $0<\alpha<2$, Zhang \cite{16} has proved

\vspace{0.3cm}{\bf Theorem 1.1(See\cite{16})}. For (\ref{6}) with
$0<\alpha<2$, there exists one odd parabolic or hyperbolic orbit
which minimizes the corresponding variational functional.

The above results are obtained for Newtonian weak force type
potentials. For the two-body problems with charges, Wu and Zhang
\cite{15} have proved the existence of hyperbolic orbits for a class
of singular Hamiltonian systems with fixed energy, they obtained the
following theorem.

\vspace{0.3cm}{\bf Theorem 1.2(See\cite{15})}\ \ {\em Suppose that
$V\in C^{1}(R^{N}\setminus\{0\},R^{1})$ satisfies

\vspace{0.3cm}{\bf $(A_{1})$}\ $V(-x)=V(x)$, $\forall x\in
R^{N}\setminus\{0\}$,

\vspace{0.3cm}{\bf $(A_{2})$}\ there is a constant $\alpha\in (0,2)$
such that
\begin{eqnarray*}
(x,\nabla V(x))= -\alpha V(x)<0\ \ \ \mbox{for}\ \ \mbox{any}\ \ \
x\in R^{N}\setminus\{0\}.
\end{eqnarray*}

Then for any $H>0$, there is at least one hyperbolic orbit for
systems (1)-(2).}

Similarly, in restricted three-body problems, we can also consider
three bodies which are charged. Suppose $e_{1}$, $e_{2}$ and $e$
represent the charges of $m_{1}$, $m_{2}$ and $m$ with
$e_{1}=e_{2}$, then the effect force between the mass points not
only obey the Newton's but also Coulomb's laws. When
$|e_{1}|=|e_{2}|$ are small enough, the first two bodies attract
each other which implies that they can move in their elliptic
orbits. The motion equation of the third mass point is
\begin{eqnarray}
 m\ddot{u}(t)+\frac{\alpha(
 m+2ee_{1})u(t)}{(|u(t)|^{2}+|r(t)|^{2})^{\frac{\alpha+2}{2}}}=0.\label{7}
\end{eqnarray}

In this model, the potential is not singular which is much different
from the Newtonian type potentials. As to the existence of periodic
orbits for non-singular Hamiltonian systems with fixed energy, there
have been many works. In 1978, Rabinowitz \cite{28} used variational
methods for strongly indefinite functionals to study the existence
of a periodic solution of a class of Hamiltonian systems on any
given regular energy hyperface. He obtained the following result.

\vspace{0.3cm}{\bf Theorem 1.3(See\cite{28})}\ \ {\em Let $H\in
C^{1}(R^{2N},R^{1})$. Suppose

\vspace{0.3cm}{\bf $(B_{1})$}\ for some $b\neq0$, $H^{-1}(b)$ is
radially homeomorphic to $S^{2N-1}$.

\vspace{0.3cm}{\bf $(B_{2})$}\ $H_{z}\neq0$, $\forall\ \zeta\in
H^{-1}(b)$.
\\

Then the Hamiltonian system
\begin{eqnarray}
 \frac{dz}{dt}=J H_{z}\label{28}
\end{eqnarray}
possesses a periodic solution on $H^{-1}(b)$, where $z=(p,q)\in
R^{2N}$, $H=H(p,q)$, $H_{z}=\left(\frac{\partial H}{\partial
p},\frac{\partial H}{\partial q}\right)$, $J=\left(\begin{array}{cc}
 0 & -I  \\
 I & 0
\end{array}
\right)_{2N\times2N}$.}

Since the pioneering work of Rabinowitz, there are many works on the
existence of periodic solutions for (\ref{28}) or second order
Hamiltonian systems. As to the unbounded orbits for non-singular
Hamiltonian systems with a fixed energy, there are only few paper
relating to this topic. In 1994, E. Serra \cite{11} has obtained the
existence of Homoclinic orbits at infinity for a class of second
order conservative systems. In his paper, he treated the systems
with zero energy and he approximated the homoclinic orbits with a
sequence of brake orbits which are obtained by variational methods.
He obtained the following theorem.

\vspace{0.3cm}{\bf Theorem 1.4(See\cite{11})}\ \ {\em Suppose that
the potential $V\in C^{2}(R^{N},R^{1})$ satisfies

\vspace{0.3cm}{\bf $(C_{1})$}\ $V(x)<0$ for all $x\in R^{N}$,

\vspace{0.3cm}{\bf $(C_{2})$}\ there exist $R_{0}>0$, $\gamma>2$
such that
\begin{eqnarray*}
V(x)=-\frac{1}{|x|^{\gamma}}+W(x),\ \ \ \forall\ |x|\geq R_{0},
\end{eqnarray*}

\vspace{0.3cm}{\bf $(C_{3})$}\
$\lim_{|x|\rightarrow+\infty}W(x)|x|^{\gamma}=0$,

\vspace{0.3cm}{\bf $(C_{4})$}\ $(x,\nabla W(x))>0$, $\forall |x|\geq
R_{0}$.
\\

Then there exists at least one solution to the problem

\begin{eqnarray}
\left\{
\begin{array}{ll}
\ddot{u}(t)+\nabla V(u(t))=0,&\mbox{for all $t\in R$,}\\
\lim_{t\rightarrow\pm\infty}|u(t)|=+\infty,&\mbox{}\\
\lim_{t\rightarrow\pm\infty}\dot{u}(t)=0.\label{8}
\end{array}
\right.
\end{eqnarray}}

\vspace{0.3cm}\textbf{Definition 1.5(See\cite{2})}\ An orbit $u(t)$
is called a parabolic orbit, if we have
\begin{eqnarray*}
|u(t)|\rightarrow +\infty,\ \ \ |\dot{u}(t)|\rightarrow 0\ \ \
\mbox{as}\ \ \ |t|\rightarrow +\infty;
\end{eqnarray*}

An orbit $u(t)$ is called a hyperbolic orbit, if we have
\begin{eqnarray*}
|u(t)|\rightarrow +\infty,\ \ \ |\dot{u}(t)|>0\ \ \ \mbox{as}\ \ \
|t|\rightarrow +\infty.
\end{eqnarray*}

Motivated by above papers, we have following theorems.

\vspace{0.3cm}{\bf Theorem 1.6}\ \ {\em Suppose $V\in
C^{1}(R^{N},R^{1})$ satisfies

\vspace{0.3cm}{\bf $(V_{1})$}\  $V(0)\geq V(-x)=V(x)>0$, for all
$x\in R^{N}$.

\vspace{0.3cm}{\bf $(V_{2})$}\ $(x,\nabla V(x))\rightarrow0,\ \ \
\mbox{as}\ \ \ |x|\rightarrow+\infty$.

\vspace{0.3cm}{\bf $(V_{3})$}\ $ V(x)\rightarrow0,\
\ \ \mbox{as}\ \ \ |x|\rightarrow+\infty$.\\

Then system (\ref{1})-(\ref{2}) possesses at least one hyperbolic
orbit for any given $H>V(0)$.}

\vspace{0.3cm}{\bf Theorem 1.7}\ \ {\em Suppose $V\in
C^{1}(R^{N},R^{1})$ satisfies $(V_{2})$, $(V_{3})$ and

\vspace{0.3cm}{\bf $(V_{4})$}\ $V(-x)=V(x)<0$, for all $x\in R^{N}$.
\\

Then systems (\ref{1})-(\ref{2}) possesses at least one hyperbolic
orbits for any $H>0$.}

\vspace{0.3cm}{\bf Remark 1}\ \ In this paper, we use the
$1/2$-antisymmetrical constraint to reduce the norm. Since the
potential in this paper has no singulary, we can also reduce the
norm on the odd-antisymmetry constrain space which is
\begin{eqnarray*}
E_{R}=\left\{q\in H^{1}|\
q(-t)=-q(t),\left|q\left(-\frac{1}{2}\right)\right|=\left|q\left(\frac{1}{2}\right)\right|=R\right\},
\end{eqnarray*}
where $H^{1}=W^{1,2}([-\frac{1}{2},\frac{1}{2}],R^{N})$. And all the
proofs are similar to this paper.

\vspace{0.3cm}{\bf Remark 2}\ \ Notice that in model (\ref{7}), if
$e$ has different sign with $e_{1}$, $e_{2}$ and $|e|$ is large
enough, the parameter $\alpha( m+2ee_{1})$ is negative, which
satisfies all conditions in Theorem 1.6. On the other hand, if $e$
has the same sign with $e_{1}$ and $e_{2}$, the parameter $\alpha(
m+2ee_{1})$ is positive, which satisfies all conditions in Theorem
1.7.

\section{Variational Settings}

\ \ \ \ \ \ Let $L^{\infty}([0,1],R^{N})$ be a space of measurable
functions from $[0,1]$ into $R^{N}$ and essentially bounded under
the following norm
\begin{eqnarray*}
\ \|q\|_{L^{\infty}([0,1],R^{N})}:=esssup\{|q(t)|:t\in[0,1]\}.
\end{eqnarray*}

Similar to A. Ambrosetti and V. Coti. Zelati in \cite{1}, we use the
$1/2$-antisymmetrical constraint to reduce the norm. The space where
we define the functional is as follow.
\begin{eqnarray*}
M_{R}=\{q\in H^{1}|\ q(t+\frac{1}{2})=-q(t),|q(0)|=|q(1)|=R\}.
\end{eqnarray*}

For any $q\in H^{1}$, we know that the following norms are
equivalent to each other
\begin{eqnarray}
&&\|q\|_{H^1}=\left(\int^{1}_{0}|\dot{q}(t)|^2dt\right)^{1/2}+\left|\int^{1}_{0}q(t)dt\right|\nonumber\\
&&\|q\|_{H^1}=\left(\int^{1}_{0}|\dot{q}(t)|^2dt\right)^{1/2}+\left(\int^{1}_{0}|q(t)|^2dt\right)^{1/2}\nonumber\\
&&\|q\|_{H^1}=\left(\int^{1}_{0}|\dot{q}(t)|^2dt\right)^{1/2}+|q(0)|.\nonumber
\end{eqnarray}

If $q\in M_{R}$, we have $\displaystyle\int^{1}_{0}q(t)dt=0$, then
by Poincar$\acute{\mbox{e}}$-Wirtinger's inequality, we obtain that
the above norms are equivalent to
\begin{eqnarray*}
\|q\|_{H^1}=\left(\int^{1}_{0}|\dot{q}(t)|^2dt\right)^{1/2}.
\end{eqnarray*}

Moreover, let $f:\ M_{R}\rightarrow R^{1}$ be the functional defined
by
\begin{eqnarray}
f(q)&=&\frac{1}{2}\int^{1}_{0}|\dot{q}(t)|^2dt\int^{1}_{0}(H-V(q(t)))dt\nonumber\\
&=& \mbox{}\frac{1}{2}\|q\|^{2}\int^{1}_{0}(H-V(q(t)))dt.\label{20}
\end{eqnarray}
Then one can easily check that $f\in C^{1}(M_{R},R^{1})$ and
\begin{eqnarray}
\ \langle
f'(q),q(t)\rangle&=&\|q\|^2\int^{1}_{0}\left(H-V(q(t))-\frac{1}{2}(
\nabla V(q(t)),q(t))\right)dt.\label{13}
\end{eqnarray}

Firstly, we prove Theorem 1.6. To prove this theorem, our way is to
approach the hyperbolic orbits with a sequence of periodic orbits
which are obtained by the minimizing theory. We need the following
lemma which is proved by A. Ambrosetti and V. Coti. Zelati in
\cite{1}.

\vspace{0.3cm}{\bf Lemma 2.1(See\cite{1})}\ {\em Let
$f(q)=\displaystyle\frac{1}{2}\int^{1}_{0}|\dot{q}(t)|^2dt\int^{1}_{0}(H-V(q(t)))dt$
and $\tilde{q}\in H^{1}$ be such that $f^{'}(\tilde{q})=0$,
$f(\tilde{q})>0$. Set
\begin{eqnarray*}
T^{2}=\frac{\displaystyle\frac{1}{2}\displaystyle\int^{1}_{0}|\dot{\tilde{q}}(t)|^{2}dt}{\displaystyle\int^{1}_{0}(H-V(\tilde{q}(t))dt}.
\end{eqnarray*}
Then $\tilde{u}(t)=\tilde{q}(t/T)$ is a non-constant $T$-periodic
solution for (\ref{1}) and (\ref{2}).}

\vspace{0.3cm}{\bf Remark 3}\ In view of the proof of Lemma 2.3 in
\cite{1}, we can see that the condition $f(\tilde{q})>0$ in Lemma
3.1 can be replaced by
$\displaystyle\int^{1}_{0}|\dot{\tilde{q}}(t)|^{2}dt>0$.

\vspace{0.3cm}{\bf Lemma 2.2(Palais\cite{13})}\ {\em Let $\sigma$ be
an orthogonal representation of a finite or compact group $G$ in the
real Hilbert space $H$ such that for any $\sigma\in G$,
\begin{eqnarray*}
f(\sigma\cdot x)=f(x),
\end{eqnarray*}
where $f\in C^{1}(H,R^{1})$. Let $S=\{x\in H|\sigma x=x,
\forall\sigma\in G\}$, then the critical point of $f$ in $S$ is also
a critical point of $f$ in $H$.}

\vspace{0.3cm}{\bf Lemma 2.3(Translation Property\cite{8})}\ {\em
Suppose that, in domain $D\subset R^{N}$, we have a solution
$\phi(t)$ for the following differential equation
\begin{eqnarray*}
x^{(n)}+F(x^{(n-1)},\cdots,x)=0,
\end{eqnarray*}
where $x^{(k)}=d^{k}x/dt^{k}$, $k=0,1,\cdots,n$, $x^{(0)}=x$. Then
$\phi(t-t_{0})$ with $t_{0}$ being a constant is also a solution.}

\section{The Proof of Theorem 1.6}

\ \ \ \ \ Firstly, we prove the existence of the approximate
solutions, then we study the limit procedure. In order to obtain the
critical points of the functional and make some estimations, we need
the following lemma.

\vspace{0.3cm}{\bf Lemma 3.1}\ {\em Suppose the conditions of
Theorem 1.6 hold, then for any  $R>0$, there exists at least one
periodic solution on $M_{R}$ for the following systems
\begin{equation}
   \ddot{q}(t)+\nabla V(q(t))=0,\ \ \ \
   \forall\
   t\in\left(-\frac{T_{R}}{2},\frac{T_{R}}{2}\right)\label{23}
\end{equation}
with
\begin{equation}
   \frac{1}{2}|\dot{q}(t)|^{2}+ V(q(t))=H,\ \ \ \ \ \
   \forall\ t\in\left(-\frac{T_{R}}{2},\frac{T_{R}}{2}\right),\label{24}
\end{equation}}
where $T_{R}$ is defined as
\begin{eqnarray}
T^{2}_{R}=\frac{\displaystyle\frac{1}{2}\displaystyle\int^{1}_{0}|\dot{q}_{R}(t)|^{2}dt}{\displaystyle\int^{1}_{0}(H-V(q_{R}(t)))dt},\label{25}
\end{eqnarray}
where $q_{R}(t)$ is the minimizer for the functional.

\vspace{0.3cm}{\bf Proof.}\ We notice that $H^{1}$ is a reflexive
Banach space and $M_{R}$ is a weakly closed subset of $H^{1}$. By
the definition of $f$, $(V_{1})$ and $H>V(0)$, we obtain that $f$ is
a functional bounded from below and
\begin{eqnarray*}
f(q)&=&\frac{1}{2}\|q\|^{2}\int^{1}_{0}(H-V(q(t)))dt\nonumber\\
&\geq& \frac{(H-V(0))}{2}\|q\|^{2}\rightarrow +\infty\ \ \
\mbox{as}\ \ \|q\|\rightarrow +\infty.
\end{eqnarray*}
Furthermore, it is easy to check that $f$ is weakly lower
semi-continuous. Then, we can see that for every $R>0$ there exists
a minimizer $q_{R}\in M_{R}$ such that
\begin{eqnarray}
f'(q_R)=0,\ \ \ \ f(q_R)=\inf_{q\in M_{R}} f(q)\geq0.\label{18}
\end{eqnarray}
It is easy to see that
$\|q_{R}\|^{2}=\int^{1}_{0}|\dot{q}_{R}(t)|^{2}dt>0$, otherwise we
deduce that $|q_{R}(t)|\equiv R>0$, on the other hand, by the
$1/2$-antisymmetry of $q_{R}$, we have $q_{R}\equiv 0$, which is a
contradiction. Then by Lemmas 2.1-2.3,
$u_{R}(t)=q_{R}(\frac{t+\frac{T_{R}}{2}}{T_{R}}):\left(-\frac{T_{R}}{2},\frac{T_{R}}{2}\right)\rightarrow
M_{R}$ is a non-constant $T_{R}$-periodic solution satisfying
(\ref{23}) and (\ref{24}). The proof of this lemma is finished.

\vspace{0.3cm}{\bf Remark 4}\ In our model, the set $M_{R}$ is a
closed set in set $H^{1}$. We minimize the functional on the set
$M_{R}$, however, we can not show that $u_{R}(t)$ solve the equation
at $\pm \frac{T_{R}}{2}$. But it is true that we do not need that
$u_{R}(t)$ is a solution at these two moments. Furthermore, we know
that $u_{R}(t)$ still has definition at $\pm
 \frac{T_{R}}{2}$ and $|u_{R}(\pm \frac{T_{R}}{2})|=R$.

Subsequently, we need to let $R\rightarrow+\infty$. But before doing
this, we need to prove $u_{R}$ can not diverge to infinity uniformly
as $R\rightarrow+\infty$, which is the following lemma.

\vspace{0.3cm}{\bf Lemma 3.2}\ {\em Suppose that
$u_{R}(t):\left[-\frac{T_{R}}{2},\frac{T_{R}}{2}\right]\rightarrow
M_{R}$ is the solution obtained in Lemma 3.1, then
$\min_{t\in\left[-\frac{T_{R}}{2},\frac{T_{R}}{2}\right]}|u_{R}(t)|$
is bounded from above. More precisely, there is a constant $M>0$
independent of $R$ such that
\begin{eqnarray*}
\min_{t\in\left[-\frac{T_{R}}{2},\frac{T_{R}}{2}\right]}|u_{R}(t)|\leq
M\ \ \ \mbox{for all}\ \ \ R>0.
\end{eqnarray*} }

\vspace{0.3cm}{\bf Proof.}\ Since $q_{R}\in M_{R}$, it is easy to
see that $u_{R}(t)=q_{R}(\frac{t+\frac{T_{R}}{2}}{T_{R}})$ satisfies
$u_{R}(-\frac{T_{R}}{2})=u_{R}(\frac{T_{R}}{2})$ and
$\dot{u}_{R}(-\frac{T_{R}}{2})=\dot{u}_{R}(\frac{T_{R}}{2})$, then
we have that
\begin{eqnarray*}
&&\left(u_{R}\left(\frac{T_{R}}{2}\right),\dot{u}_{R}\left(\frac{T_{R}}{2}\right)\right)
-\left(u_{R}\left(-\frac{T_{R}}{2}\right),\dot{u}_{R}\left(-\frac{T_{R}}{2}\right)\right)\nonumber\\
&=& \mbox {}
\int^{\frac{T_{R}}{2}}_{-\frac{T_{R}}{2}}\frac{d}{dt}(u_{R}(t),\dot{u}_{R}(t))dt\nonumber\\
&=& \mbox {} \int^{\frac{T_{R}}{2}}_{-\frac{T_{R}}{2}}
(|\dot{u}_{R}(t)|^{2}+(u_{R}(t),\ddot{u}_{R}(t)))dt\nonumber\\
&=& \mbox {}
\int^{\frac{T_{R}}{2}}_{-\frac{T_{R}}{2}}2(H-V(u_{R}(t)))-(\nabla
V(u_{R}(t)),u_{R}(t))dt.
\end{eqnarray*}
Then we obtain that
\begin{eqnarray*}
\int^{\frac{T_{R}}{2}}_{-\frac{T_{R}}{2}}2H-(2V(u_{R}(t))+(\nabla
V(u_{R}(t)),u_{R}(t)))dt=0.
\end{eqnarray*}
There are two cases needed to be discussed.

{\bf Case 1.}\ $2H-(2V(u_{R}(t))+(\nabla
V(u_{R}(t)),u_{R}(t)))\equiv0$, which implies that
\begin{eqnarray*}
2H&=&2V(u_{R}(t))+(\nabla V(u_{R}(t)),u_{R}(t)),\ \ \ \mbox{a.e.} \
t\in\left[-\frac{T_{R}}{2},\frac{T_{R}}{2}\right].
\end{eqnarray*}
Hypotheses $(V_{2})$, $(V_{3})$ imply that there exists a constant
$M_{1}>0$ independent of $R$ such that
\begin{eqnarray*}
\min_{t\in\left[-\frac{T_{R}}{2},\frac{T_{R}}{2}\right]}|u_{R}(t)|\leq
M_{1}.
\end{eqnarray*}

{\bf Case 2.}\ $2(H-V(u_{R}(t)))-(\nabla V(u_{R}(t)),u_{R}(t))$
changes sign in $\left[-\frac{T_{R}}{2},\frac{T_{R}}{2}\right]$.
Then there exists
$t_{0}\in\left[-\frac{T_{R}}{2},\frac{T_{R}}{2}\right]$ such that
\begin{eqnarray*}
2H-(2V(u_{R}(t_{0}))+(\nabla V(u_{R}(t_{0})),u_{R}(t_{0})))<0,
\end{eqnarray*}
which implies that
\begin{eqnarray*}
2H&<&2V(u_{R}(t_{0}))+(\nabla V(u_{R}(t_{0})),u_{R}(t_{0})).
\end{eqnarray*}
It follows from $H>0$ and hypotheses $(V_{2})$, $(V_{3})$ that there
exists a constant $M_{2}>0$ independent of $R$ such that
\begin{eqnarray*}
\min_{t\in\left[-\frac{T_{R}}{2},\frac{T_{R}}{2}\right]}|u_{R}(t)|\leq
M_{2}.
\end{eqnarray*}
Then the proof is completed.

\vspace{0.3cm}{\bf Lemma 3.3}\ {\em Suppose that $R>M$ and
$u_{R}(t)$ is the solution for $(\ref{23})-(\ref{24})$ obtained in
Lemma 3.1, where $M$ is from Lemma 3.2. Set
\begin{eqnarray*}
t_{+}=\sup\left\{t\in\left[-\frac{T_{R}}{2},\frac{T_{R}}{2}\right]|\left|u_{R}(t)\right|\leq
L\right\}
\end{eqnarray*}
and
\begin{eqnarray*}t_{-}=\inf\left\{t\in\left[-\frac{T_{R}}{2},\frac{T_{R}}{2}\right]|\left|u_{R}(t)\right|\leq
L\right\}
\end{eqnarray*}
where $L$ is a constant independent of $R$ such that $M<L<R$. Then
we have that}
\begin{eqnarray*}
\frac{T_{R}}{2}-t_{+}\rightarrow+\infty,\ \ \
t_{-}+\frac{T_{R}}{2}\rightarrow+\infty\ \ \ \mbox{as}\ \
R\rightarrow+\infty.
\end{eqnarray*}

\vspace{0.3cm}{\bf Proof.}\ By the definition of $u_{R}(t)$ we have
that
\begin{eqnarray*}
\left|u_{R}\left(-\frac{T_{R}}{2}\right)\right|=\left|u_{R}\left(\frac{T_{R}}{2}\right)\right|=R.
\end{eqnarray*}
Then, by $(V_{1})$ and the definitions of $t_{+}$, we have
\begin{eqnarray}
\int^{\frac{T_{R}}{2}}_{t_{+}}\sqrt{H-V(u_{R}(t))}|\dot{u}_{R}(t)|dt&\geq&\int^{\frac{T_{R}}{2}}_{t_{+}}\sqrt{H-V(0)}|\dot{u}_{R}(t)|dt\nonumber\\
&\geq& \mbox{}
\sqrt{H-V(0)}\int^{\frac{T_{R}}{2}}_{t_{+}}|\dot{u}_{R}(t)|dt\nonumber\\
&\geq& \mbox{}
\sqrt{H-V(0)}\left|\int^{\frac{T_{R}}{2}}_{t_{+}}\dot{u}_{R}(t)dt\right|\nonumber\\
&\geq& \mbox{}\sqrt{H-V(0)}(R-L).\label{3}
\end{eqnarray}
It follows from Lemma 3.1 and $(V_{1})$ that
\begin{eqnarray*}
\int^{\frac{T_{R}}{2}}_{t_{+}}\sqrt{H-V(u_{R}(t))}|\dot{u}_{R}(t)|dt&=&\sqrt{2}\int^{\frac{T_{R}}{2}}_{t_{+}}\sqrt{H-V(u_{R}(t))}\sqrt{H-V(u_{R}(t))}dt\nonumber\\
&\leq& \mbox{} \sqrt{2}H\left(\frac{T_{R}}{2}-t_{+}\right)
\end{eqnarray*}
Combining (\ref{3}) with the  above estimation, we obtain that
\begin{eqnarray*}
\sqrt{H-V(0)}(R-L)\leq\sqrt{2}H\left(\frac{T_{R}}{2}-t_{+}\right).
\end{eqnarray*}
Then we have
\begin{eqnarray*}
\frac{T_{R}}{2}-t_{+}\rightarrow+\infty,\ \ \ \mbox{as}\ \
R\rightarrow+\infty.
\end{eqnarray*}
The limit for $t_{-}+\frac{T_{R}}{2}$ can be obtained in the similar
way. The proof is completed.

\vspace{0.3cm}{\bf The Limit Procedure}\ Subsequently, we set that
\begin{eqnarray*}
t^{*}=\inf\left\{t\in\left[-\frac{T_{R}}{2},\frac{T_{R}}{2}\right]||u_{R}(t)|=M\right\}
\end{eqnarray*}
and
\begin{eqnarray*}
u_{R}^{*}(t)=u_{R}(t-t^{*})
\end{eqnarray*}
Since $L>M$, we can deduce that $t_{+}\geq t^{*}\geq t_{-}$, which
implies that
\begin{eqnarray*}
-\frac{T_{R}}{2}+t^{*}\rightarrow-\infty,\ \ \mbox{}\ \
\frac{T_{R}}{2}+t^{*}\rightarrow+\infty\ \ \mbox{as}\ \
R\rightarrow\infty.
\end{eqnarray*}

Then it follows from $(13)$ that
\begin{eqnarray*}
   \frac{1}{2}|\dot{u}^{*}_{R}(t)|^{2}+ V(u^{*}_{R}(t))=H,\ \ \
   \forall\ t\in\left(-\frac{T_{R}}{2}+t^{*},\frac{T_{R}}{2}+t^{*}\right).
\end{eqnarray*}
which implies that
\begin{eqnarray*}
|\dot{u}^{*}_{R}(t)|^{2}=2(H-V(u^{*}_{R}(t))),\ \ \
  \forall\ t\in\left(-\frac{T_{R}}{2}+t^{*},\frac{T_{R}}{2}+t^{*}\right).
\end{eqnarray*}
By $(V_{3})$ and $V\in C^{1}(R^{N}, R^{1})$, we can deduce that
there exists a constant $M_{4}>0$ independent of $R$ such that
\begin{eqnarray*}
 |V(u^{*}_{R}(t))|\leq M_{4}\ \ \ \mbox{for all}\ \
 t\in\left(-\frac{T_{R}}{2}+t^{*},\frac{T_{R}}{2}+t^{*}\right).
\end{eqnarray*}
Then there is a constant $M_{5}$ independent of $R$ such that
\begin{eqnarray*}
 |\dot{u}^{*}_{R}(t)|\leq M_{5}\ \ \ \mbox{for all}\ \
 t\in\left(-\frac{T_{R}}{2}+t^{*},\frac{T_{R}}{2}+t^{*}\right).
\end{eqnarray*}
which implies that
\begin{eqnarray*}
\ |u_{R}^{*}(t_{1})-u_{R}^{*}(t_{2})| \leq
\left|\int^{t_{1}}_{t_{2}}\dot{u}_{R}^{*}(s)ds\right| \leq \mbox{}
\int^{t_{1}}_{t_{2}}\left|\dot{u}_{R}^{*}(s)\right|ds \leq
M_{5}|t_{1}-t_{2}|
\end{eqnarray*}
for each $R>0$ and $t_{1}, t_{2} \in
\left(-\frac{T_{R}}{2}+t^{*},\frac{T_{R}}{2}+t^{*}\right)$, which
shows $\{u_{R}^{*}\}$ is equicontinuous. Then there is a subsequence
$\{u_{R}^{*}\}_{R>0}$ converging to $u_{\infty}$ in
$C_{loc}(R^{1},R^{N})$. Then there exists a function $u_{\infty}(t)$
such that
\begin{eqnarray*}
&&(\mbox{i})\ u_{R}^{*}(t)\rightarrow u_{\infty}(t)\ \mbox{in}\ C_{loc}(R^{1},R^{N})\nonumber\\
&&(\mbox{ii})|u_{\infty}(t)|\rightarrow +\infty\ \mbox{as}\
|t|\rightarrow+\infty
\end{eqnarray*}
and $u_{\infty}(t)$ satisfies systems $(1)-(2)$.

From the above lemmas, we have proved there is at least one
hyperbolic solution for $(1)-(2)$ with $H>0$. We finish the proof of
Theorem 1.6.

\section{The Proof of Theorem 1.7}

Since the potential in Theorem 1.7 is negative and of $C^{1}$ class,
the proof of this theorem is more simple. Similar to the proof of
Theorem 1.6, we consider the functional (\ref{20}) on $M_{R}$ which
is $f:\ M_{R}\rightarrow R$.

\vspace{0.3cm}{\bf Lemma 4.1}\ {\em Suppose the conditions of
Theorem 1.5 hold, then for any  $R>0$, there exists at least one
periodic solution on $M_{R}$ for the following systems
\begin{equation}
   \ddot{q}(t)+\nabla V(q(t))=0,\ \ \ \
   \forall\ t\in\left(-\frac{T_{R}}{2},\frac{T_{R}}{2}\right)\label{16}
\end{equation}
with
\begin{equation}
   \frac{1}{2}|\dot{q}(t)|^{2}+ V(q(t))=H,\ \ \ \ \ \
   \forall\ t\in\left(-\frac{T_{R}}{2},\frac{T_{R}}{2}\right).\label{17}
\end{equation}}

\vspace{0.3cm}{\bf Proof.}\ We notice that $H^{1}$ is a reflexive
Banach space and $M_{R}$ is a weakly closed subset of $H^{1}$. Since
$H>0$, we obtain that
\begin{eqnarray*}
f(q)&=&\frac{1}{2}\|q\|^{2}\int^{1}_{0}(H-V(q(t))dt\geq\frac{H}{2}\|q\|^{2},
\end{eqnarray*}
which implies that $f$ is a functional bounded from below,
furthermore, it is easy to check that $f$ is weakly lower
semi-continuous and
\begin{eqnarray*}
f(q)\rightarrow +\infty\ \ \ \mbox{as}\ \ \|q\|\rightarrow +\infty.
\end{eqnarray*}

Then, we conclude that for every $R>0$ there exists a minimizer
$Q_{R}\in M_{R}$ such that
\begin{eqnarray*}
f'(Q_R)=0,\ \ \ \ f(Q_R)=\inf_{q\in M_{R}} f(q)>0.
\end{eqnarray*}

It is easy to see that
$\|Q_{R}\|^{2}=\int^{1}_{0}|\dot{Q}_{R}(t)|^{2}dt>0$, otherwise we
deduce that $|Q_{R}(t)|\equiv R$, on the other hand, by the
$1/2$-antisymmetry of $Q_{R}$, we have $Q_{R}\equiv 0$, which is a
contradiction. This implies that $f(Q_R)>0$. Then let
\begin{eqnarray}
T^{2}_{R}=\frac{\displaystyle\frac{1}{2}\displaystyle\int^{1}_{0}|\dot{Q}_{R}(t)|^{2}dt}{\displaystyle\int^{1}_{0}(H-V(Q_{R}(t)))dt},\label{29}
\end{eqnarray}
then by Lemmas 2.1-2.3,
$U_{R}(t)=Q_{R}(\frac{t+\frac{T_{R}}{2}}{T_{R}}):\left(-\frac{T_{R}}{2},\frac{T_{R}}{2}\right)\rightarrow
M_{R}$ is a non-constant $T_{R}$-periodic solution satisfying
(\ref{16}) and (\ref{17}). The proof of this lemma is finished.

Subsequently, we need to show that $U_{R}(t)$ can not diverge to
infinity uniformly as $R\rightarrow+\infty$. Moreover, we prove the
following lemma.

\vspace{0.3cm}{\bf Lemma 4.2}\ {\em Suppose that
$U_{R}(t):\left[-\frac{T_{R}}{2},\frac{T_{R}}{2}\right]\rightarrow
M_{R}$ is the solution obtained in Lemma 4.1, then
$\min_{t\in\left[-\frac{T_{R}}{2},\frac{T_{R}}{2}\right]}|U_{R}(t)|$
is bounded from above. More precisely, there is a constant $M'>0$
independent of $R$ such that
\begin{eqnarray*}
\min_{t\in\left[-\frac{T_{R}}{2},\frac{T_{R}}{2}\right]}|U_{R}(t)|\leq
M'\ \ \ \mbox{for all}\ \ \ R>0.
\end{eqnarray*} }
The proof of this lemma is same with that of Lemma 3.2.

\vspace{0.3cm}{\bf Lemma 4.3}\ {\em Suppose that $R>M'$, where $M'$
is defined in Lemma 4.2 and $U_{R}(t)$ is the solution for
(\ref{16}) and (\ref{17}) obtained in Lemma 4.1. Set
\begin{eqnarray*}
t_{+}=\sup\left\{t\in\left[-\frac{T_{R}}{2},\frac{T_{R}}{2}\right]|\left|U_{R}(t)\right|\leq
l\right\}
\end{eqnarray*}
and
\begin{eqnarray*}t_{-}=\inf\left\{t\in\left[-\frac{T_{R}}{2},\frac{T_{R}}{2}\right]|\left|U_{R}(t)\right|\leq
l\right\}
\end{eqnarray*}
where $l$ is a constant independent of $R$ such that $M'<l<R$. Then
we have that}
\begin{eqnarray*}
\frac{T_{R}}{2}-t_{+}\rightarrow+\infty,\ \ \
t_{-}+\frac{T_{R}}{2}\rightarrow+\infty\ \ \ \mbox{as}\ \
R\rightarrow+\infty.
\end{eqnarray*}

\vspace{0.3cm}{\bf Proof.}\ By the definition of $U_{R}(t)$ we have
that
\begin{eqnarray*}
\left|U_{R}\left(-\frac{T_{R}}{2}\right)\right|=\left|U_{R}\left(\frac{T_{R}}{2}\right)\right|=R.
\end{eqnarray*}
Then, by $(V_{4})$ and the definitions of $t_{+}$ and $t_{-}$, we
have
\begin{eqnarray}
\int^{\frac{T_{R}}{2}}_{t_{+}}\sqrt{H-V(U_{R}(t))}|\dot{U}_{R}(t)|dt&\geq&\sqrt{H}\int^{\frac{T_{R}}{2}}_{t_{+}}|\dot{U}_{R}(t)|dt\nonumber\\
&\geq& \mbox{}
\sqrt{H}\left|\int^{\frac{T_{R}}{2}}_{t_{+}}\dot{U}_{R}(t)dt\right|\geq\sqrt{H}(R-l)\label{26}
\end{eqnarray}
and
\begin{eqnarray}
\int^{t_{-}}_{-\frac{T_{R}}{2}}\sqrt{H-V(U_{R}(t))}|\dot{U}_{R}(t)|dt&\geq&\sqrt{H}\int^{t_{-}}_{-\frac{T_{R}}{2}}|\dot{U}_{R}(t)|dt
\nonumber\\
&\geq& \mbox{}
\sqrt{H}\left|\int^{t_{-}}_{-\frac{T_{R}}{2}}\dot{U}_{R}(t)dt\right|\geq\sqrt{H}(R-l)\label{27}.
\end{eqnarray}
Since $V\in C^{1}(R^{N}, R^{1})$, it follows from $(V_{3})$, that
there exists a constant $M_6>0$ independent of $R$ such that
\begin{eqnarray*}
|V(U_{R}(t))|\leq M_{6}\ \ \ \mbox{for}\ \ \mbox{all}\ \
t\in\left[-\frac{T_{R}}{2},\frac{T_{R}}{2}\right],
\end{eqnarray*}
which implies that
\begin{eqnarray*}
\int^{\frac{T_{R}}{2}}_{t_{+}}\sqrt{H-V(U_{R}(t))}|\dot{U}_{R}(t)|dt=\sqrt{2}\int^{\frac{T_{R}}{2}}_{t_{+}}(H-V(U_{R}(t)))dt\leq
\sqrt{2}(H+M_{6})\left(\frac{T_{R}}{2}-t_{+}\right).
\end{eqnarray*}
Combining (\ref{26}) with the  above estimates, we obtain that
\begin{eqnarray*}
\sqrt{H}(R-L)\leq\sqrt{2}(H+M_{6})\left(\frac{T_{R}}{2}-t_{+}\right).
\end{eqnarray*}
Then we have
\begin{eqnarray*}
\frac{T_{R}}{2}-t_{+}\rightarrow+\infty,\ \ \ \mbox{as}\ \
R\rightarrow+\infty.
\end{eqnarray*}
The limit for $t_{-}+\frac{T_{R}}{2}$ can be obtained in the similar
way. The proof is completed.

The following limit procedure is similar to the proof in Theorem
1.6. $\Box$

\end{document}